\documentclass[aip,prf,reprint, amsmath,amssymb, onecolumn]{revtex4-1}

\usepackage[pdftex]{graphicx}
\usepackage{dcolumn}
\usepackage{bm}
\usepackage[T1]{fontenc}
\usepackage[utf8]{inputenc}
\usepackage{natbib}
\usepackage{hyperref}
\usepackage{color}
\usepackage{url}

\usepackage{amssymb,amsmath,amsfonts,graphicx}
\usepackage{comment}
\usepackage{subfigure}
\usepackage{xcolor}
\usepackage[english,francais]{babel}

\definecolor{OrangePastel}{RGB}{255,200,31}
\definecolor{GreenPastel}{RGB}{33,219,77}
\definecolor{VioletPastel}{RGB}{200,175,242}
\definecolor{RedPastel}{RGB}{255,125,82}

\definecolor{GreenDarkPastel}{RGB}{33,150,77}
\definecolor{GreenLightPastel}{RGB}{33,255,77}

\definecolor{GreenBabethPastel}{RGB}{209,255,108}
\definecolor{BlueBabethPastel}{RGB}{119,207,255}

\definecolor{red}{rgb}{1,0,0}
\definecolor{blue}{rgb}{0,0,1}

\newcommand{\beq}{\begin{equation}}
\newcommand{\eeq}{  \end{equation}}

\begin{document}


\title{Size of the top jet drop produced by bubble bursting}

\author{Elisabeth Ghabache}
\affiliation{Sorbonne Universit\'es, Universit\'e Pierre et Marie Curie and Centre National de la Recherche 
Scientifique, Unit\'e Mixte de Recherche 7190, Institut Jean Le Rond d'Alembert, 4 
Place Jussieu, F-75005 Paris, France}

\author{Thomas S\'eon}\email[Corresponding author: ]{thomas.seon@gmailcom}
\affiliation{Sorbonne Universit\'es, Universit\'e Pierre et Marie Curie and Centre National de la Recherche 
Scientifique, Unit\'e Mixte de Recherche 7190, Institut Jean Le Rond d'Alembert, 4 
Place Jussieu, F-75005 Paris, France}

\date{\today }

\begin{abstract}
As a bubble bursts at a liquid-air interface, a tiny liquid jet rises and can release the so-called \textit{jet drops}. In this paper, the size of the top jet drop produced by a bubble bursting is investigated experimentally. We determine, and discuss, the first scaling law enabling the determination of the top jet drop size as a function of the corresponding mother bubble radius and the liquid properties (viscosity, surface tension, density), along with its regime of existence. Furthermore, in the aim of decoupling experimentally the effects of bubble collapse and jet dynamics on the drop detachment, we propose a new scaling providing the top drop size only as a function of the jet velocity and liquid parameters. In particular, this allows us to untangle the intricate roles of viscosity, gravity and surface tension in the \textit{end-pinching} of the bubble bursting jet. 
\end{abstract} 


\maketitle

After a bubble rises in ocean, it reaches the surface and the thin film which separates the bubble from the atmosphere, the bubble cap, drains and ruptures under the effect of gravity \cite{Veron2015}. From then, two events in a row are producing droplets : the film shattering which expels $O(10-100)$ of small \textit{film drops} \cite{Spiel1998, Lhuissier2012}, and the capillary collapse of the remaining cavity which shoots up a central jet, that becomes unstable and breaks up into several larger \textit{jet drops} \cite{Veron2015}. Most of film drops are less than 1 $\mu$m  in radius while jet drops span in the range from 2 to 500  $\mu$m \cite{Leeuw2011}.
Sea spray is largely attributed to an estimated 10$^{18}$ to 10$^{20}$ bubbles that burst every second across the oceans \cite{Stuhlman1932, Woodcock1953, Wu1981, Spiel1997, Lewis2004, ODowd2007, Keene2007, Fuentes2010}. 
Main consequences of this aerosol are, a global emission of about 10$^{12}$ to 10$^{14}$ kg per year of sea salt \cite{Textor2006} and heat and momentum transfer with the atmosphere through direct exchange \cite{Andreas2008, Veron2015}.

\begin{figure*}[ht]
\centering
\noindent\includegraphics[width=\hsize]{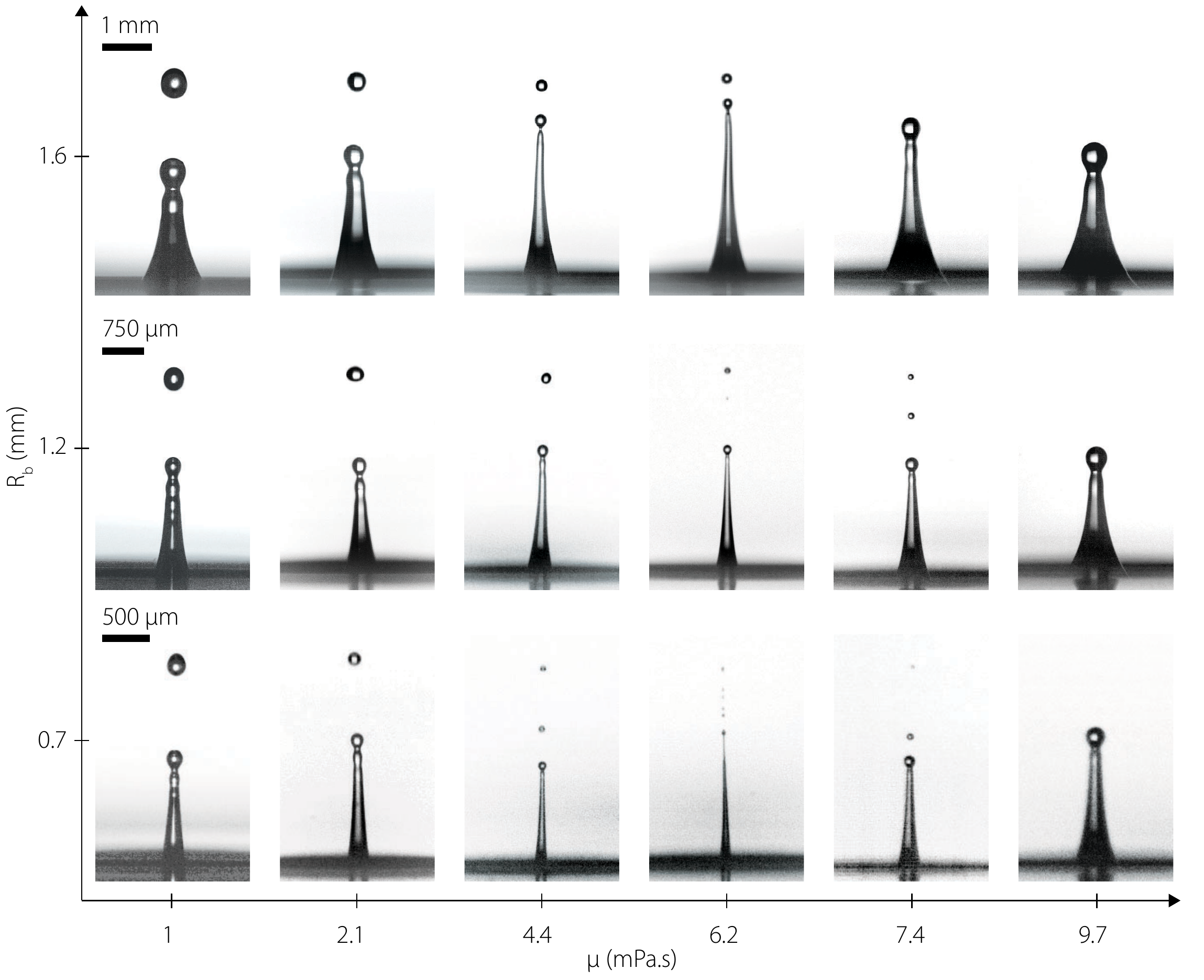}
\caption{
Snapshot of a typical jetting event following a bubble bursting at a free surface. The jets and drops shape is displayed for three mother bubble radii, reported on the y-axis, bursting in water and five water-glycerol mixtures of viscosity indicated on the x-axis. For those six solutions, the surface tension is almost constant (ranging from 64 to 72 mN.m$^{-1}$) so that one mainly observes in this figure the effect of changing viscosity. top drop size decreases with bubble radius and increasing liquid viscosity. The biggest drop, on the top left corner of this diagram, is about 400 $\mu$m radius and the smallest ($R_b=0.7$ mm and $\mu =$ 7.4 mPa.s) reached 20 $\mu$m. The scale bar is showed on the top left corner of each bubble radius and is the same the whole row.
} 
\label{fig:sequence_jet}
\end{figure*}

On a smaller scale, situation found in glasses of champagne and sparkling wines is comparable with, however a main difference :  liquid properties of a hydro-alcoholic solution are different, the surface tension is lower ($\gamma = $ 48 mN.m$^{-1}$) and, because champagne is always served at low temperature, liquid viscosity ranges from $\mu = $1.6 to 3.6 mPa.s during tasting \cite{Liger-Belair2008}. Subsequent consequences of these different properties include : almost no film drops  are produced above a glass of champagne \cite{Ghabache2016} and dynamics of jet drops is strongly modified by liquid parameters \cite{Ghabache2014}.
Furthermore, we showed in a recent study~\cite{Ghabache2015} that the top jet drops, which bound the edge of the aerosol cloud, highly dominate the evaporation process as they are faster and usually bigger than the others or with a comparable size.
In this paper, the size of the \textit{top jet drop} produced by bubble bursting is investigated as a function of the mother bubble size \cite{Hayami1958, Spiel1994, Blanchard1989} and the liquid properties (see Fig.~\ref{fig:sequence_jet}).

An infinite cylinder of liquid at rest, subjected to the influence of surface tension, will break up into a number of individual droplets through the so-called Rayleigh-Plateau instability \cite{Savart1833, Plateau1873, Rayleigh1878}. The bubble bursting jets, depicted in Fig.~\ref{fig:sequence_jet}, are finite and do not break as a consequence of Rayleigh-Plateau instability. Instead, the breakup takes place at the jet tip and detaches one drop at a time. This mechanism, called \textit{end-pinching}, consists of a competition between the capillary retraction of the jet tip, shaping a blob \cite{Taylor1959, Culick1960, Keller1983, Keller1995}, and a pressure-driven flow from the cylindrical jet toward the bulbous end. This leads to the development of a neck, where the jet joins the blob, and thus to the drop detachment via a capillary pinch off process. This mechanism has been first described in the context of a strongly deformed viscous drop \cite{Taylor1934, Stone1986, Stone1989} and later for a free liquid filament of arbitrary viscosity \cite{Schulkes1996, Notz2004, Castrejon-Pita2012, Hoepffner2013}. 
This end-pinching capillary breakup of liquid jets is important in several industrial contexts \cite{Eggers2008}, especially because of the broad range of applications of inkjet printing technology. Indeed, it enables accurate drop deposition of liquids, and includes production of organic thin-film transistors, Liquid Crystal Displays (LCD), 
fuel or solar cells, Printed Circuit Boards (PCB),
dispensing of DNA and protein substances, or even
fabrication of living tissue \cite{Williams2006, Singh2010, Wijshoff2010, Bos2014}.
Recently, the end-pinching of a stretched inertially driven jet shooting up after a cavity collapse has been described theoretically and numerically \cite{Gordillo2010}. These stretched jets are found in many situations \cite{Eggers2008, Antkowiak2007, Seon2012, Ghabache2014a}, in particular bubble bursting, and they all have similar properties. 


Our paper aims thus at contributing to the understanding and characterizing of the end-pinching of such stretched jets. This will be realized through the experimental characterization of the size of the top jet drop and its variations with respect to its natural control parameters, when jet droplet is produced by a single bubble bursting at a calm liquid surface. Scaling laws of the drop diameter along with their regime of existence will be determined and discussed.



\begin{figure}[ht]
\centering
\noindent\includegraphics[width=\hsize]{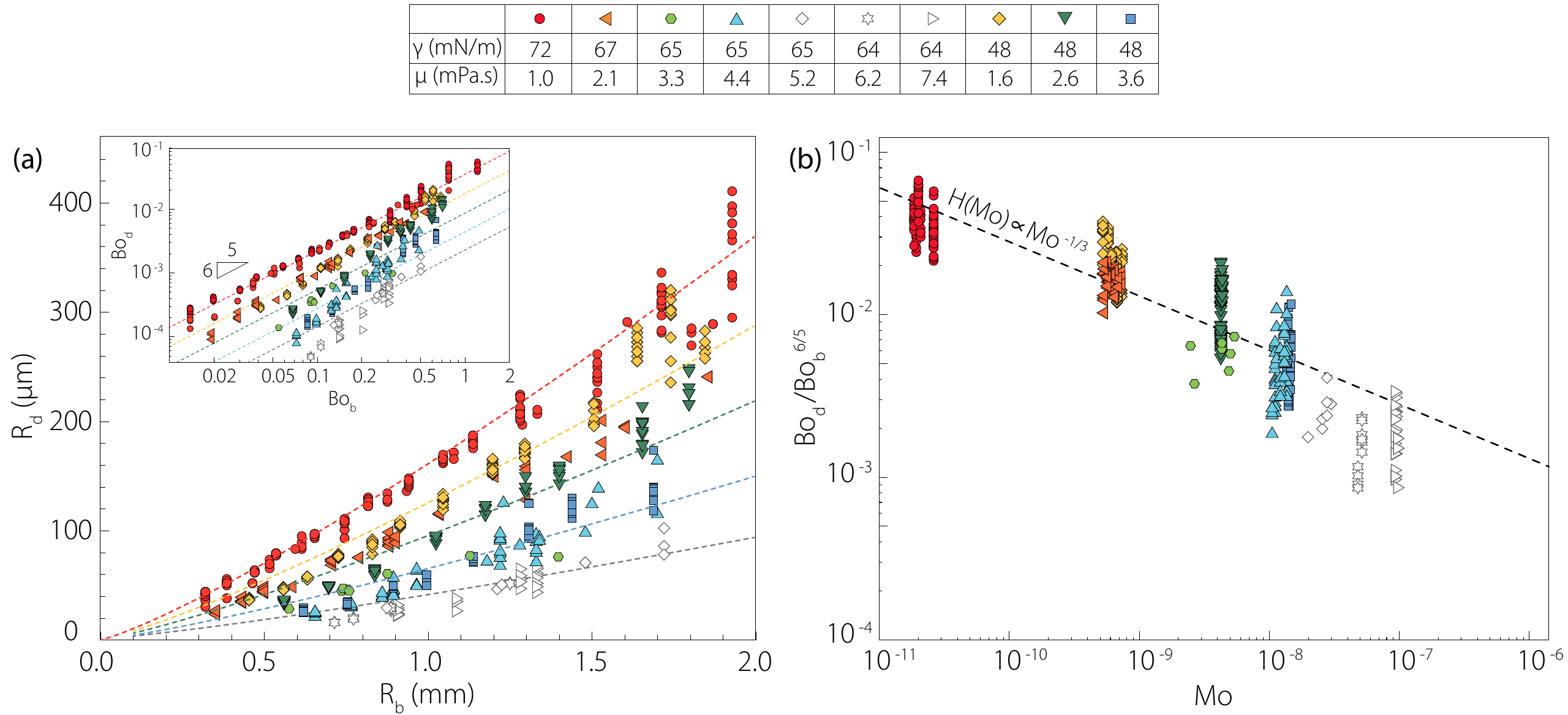}
\caption{(a) top drop radius as a function of the mother bubble radius for bubble bursting in liquids with different surface tension and viscosity. The parameters of these liquids associated to the corresponding symboles are summarized in the table above the graph. In the inset, the Bond number built on the drop radius is plotted as a function of the Bond number built on the bubble radius for the same liquids. The dashed lines represent $R_d\propto R_b^{6/5}$ in the graph and $Bo_d \propto Bo_b^{6/5}$ in the inset. (b)  $\text{Bo}_d / \text{Bo}_b^{6/5}$  as a function of the Morton number. The dashed line fits the experimental data plotted with closed symbols, up to Mo $\sim 10^{-8}$, following the trend $\text{Bo}_d / \text{Bo}_b^{6/5} \propto \text{Mo}^{-1/3}$.
}
\label{Bod_vs_Bob}
\end{figure}

The experiment consists in releasing a gas bubble from a submerged needle in a liquid and 
recording the upward jet and released drops after the bubble bursts at the free surface.
Different needle diameters allow us to create bubbles with various radii ($R_b$) ranging from 0.3 to 2~mm.
The liquids used include eleven solutions of water-glycerol-ethanol mixtures of viscosity in the range $\mu$ = 1 - 9.7 mPa.s, surface tension $\gamma$ = 48 - 72 mN.m$^{-1}$, and density $\rho$~=~980 - 1140~kg.m$^{-3}$. The surface tension and viscosity of each solution is presented along with the corresponding symbol in the table at the top of Fig.~\ref{Bod_vs_Bob}.  
The jet dynamics is analyzed through extreme close-up ultra-fast imagery, using a digital high-speed camera 
(Photron SA-5). 
Macro lenses and extension rings allow us to record with a definition reaching 5 $\mu$m per pixel. 



Figure~\ref{fig:sequence_jet} presents the jet and released drop shape following bubble bursting. 
In the cases where no drop detaches the jets are displayed at their maximum height.
On the x and y-axis the jets and drops shape is represented, respectively, for six different liquid viscosities and three different mother bubble radii.
It is clear on this diagram that, independently of the viscosity, the bigger the bubble the bigger the top drop. This intuitive result has been observed in water in various previous studies \cite{Hayami1958, Blanchard1989, Spiel1994}. Although mentioned in a earlier paper \cite{Ghabache2014}, the variation with viscosity is much more unexpected. Indeed, irrespectively of the bubble radius in the range considered here, the top drop shrinks as viscosity is increased, and seems to reach a minimum for a liquid viscosity around 6-7 mPa.s here. For higher viscosities, no drop is detached, in accordance with previous study \cite{Walls2015}. This decrease of the drop radius with viscosity is surprising, in particular because 
the Ohnesorge number based on the drop radius, namely Oh$ = \mu/\sqrt{\rho R_d \gamma}$, which compares the effect of viscosity and capillarity, is included between $10^{-1}$ and $10^{-3}$ and is consequently always lower than 1. 
This therefore suggests that  viscous effects should be neglected in the description of jet breakup, as done in similar cases \cite{Gordillo2010}. We will see further down why, in this particular case of bubble bursting jet, the liquid viscosity has such a strong influence.


We now plot, in Fig.~\ref{Bod_vs_Bob} (a), the variation of the top drop radius $R_d$ as a function of the mother bubble radius $R_b$ for different values of the liquid parameters ($\mu$, $\gamma$ and $\rho$) indicated in the table above. This quantifies our previous observation of drop shrinking with decreasing bubble radius and increasing liquid viscosity, from 400 $\mu$m to 20 $\mu$m for the solutions plotted here. 
We also observe that the same drop shrinking occurs when surface tension is decreased. 
Moreover, it appears that, regardless of the liquid parameters considered in this graph, the drop size increases with bubble radius following roughly the same variation for all the curves : $R_d \propto R_b^{6/5}$, shown with dashed lines on the graph.
Note that the historical law, proposed for the top jet drop radius produced by bubble bursting in water, that predicts a drop radius being the tenth of the bubble radius ($R_d=R_b/10$) \cite{Kientzler1954}, is only valid for bubble radii smaller than five hundred micrometers. 
More accurate laws have been written ever since. In particular, when $R_b  \geq 0.1$~mm, the relationship  $R_d = 0.075 R_b^{1.3}$ has been proposed, with radii expressed in millimeters \cite{Lewis2004, Massel2007}.
This variation is very closed to the one we find.

Thanks to our experimental relationships between $R_d$ and $R_b$, we are now able to write a more universal scaling law, taking into account the liquid parameters. 
It is clear that top drop size depends on bubble radius $R_b$, liquid viscosity $\mu$ and we assume that surface tension $\gamma$, density $\rho$ and gravity $g$ might also influence its selection, yielding : 
$$R_d=\Pi(R_b, \rho, \mu, \gamma, g).$$ 
Using dimensional arguments \cite{Buckingham1914}, this equation becomes a relation between three dimensionless numbers fully describing the top drop size selection : 
$$\text{Bo}_d = F(\text{Bo}_b, \text{Mo})$$
where the Bond numbers Bo$_d = \rho g R_d^2/\gamma$ and Bo$_b = \rho g R_b^2/\gamma$ compare the effect of gravity and capillarity on the top drop and the initial bubble respectively, and the Morton number Mo$ = g\mu^4/\rho \gamma^3$ only depends on the fluid properties and is, in particular, independent of the bubble radius $R_b$. 
On the inset of Fig.~\ref{Bod_vs_Bob} (a) the variation of the top drop Bond number is plotted as a function of the mother bubble Bond number  for the same solutions as in Fig.~\ref{Bod_vs_Bob} (a). The variation $\text{Bo}_d \propto  \text{Bo}_b^{6/5}$ is also plotted with dashed lines. This power law, independent of the liquid parameters, still works reasonably well, allowing us to write the following scaling law :
\beq
\text{Bo}_d = \text{Bo}_b^{6/5} H(\text{Mo}).
\label{BoLaw}
\eeq
In the aim of estimating the dependance of the drop size with the liquid properties, namely H(Mo), $\text{Bo}_d / \text{Bo}_b^{6/5}$ is plotted as a function of the Morton number on Fig.~\ref{Bod_vs_Bob} (b). We observe that the data with closed symbols gather along a line, up to Mo $\simeq 10^{-8}$, corresponding to a viscosity $\mu = 5.2$~mPa.s for a water-glycerol mixture. 
This line is properly fitted by H(Mo) = $\mathcal{A}$ Mo$^{-1/3}$ with $\mathcal{A}$ = 1.1~10$^{-5}$.
Therefore, in this regime, ranging on around three decades in Morton number, we established a scaling law for the top drop size as a function of the bubble radius and liquid parameters, in the context of bubble bursting : 
\beq
\text{Bo}_d = \mathcal{A}  \text{Bo}_b^{6/5} \text{Mo}^{-1/3}.
\label{BoLaw}
\eeq
This result is essential because the bubble radius and the liquid parameters are the natural experimental parameters for bursting bubble aerosol measurement. In particular, the size distribution of bubbles is know in ocean \cite{Deane2002} and can even be controlled in a glass of champagne \cite{Liger-Belair2013}.


However, under this form, Eq.\ref{BoLaw} is delicate to interpret, in particular, the confusing role of viscosity, that is expected to be negligible (Oh $\ll$ 1). In addition, this scaling law contains substantial experimental data scattering due to an accumulation of variability, when the jet is created and when the drop is detached. In the following, we wish, therefore, to express the drop radius as a function of only jet parameters, typically by disposing of the bubble radius.

When a bubble collapses, a jet is formed with a given shape, tip velocity, local strain rate etc.
In this regime, where Mo $\lesssim 10^{-8}$, the decrease of the drop size with Morton number comes along with a thinning down of the whole jet and an increase of the jet velocity. This has been largely discussed in a previous study \cite{Ghabache2014} and the corresponding scaling law, for the jet velocity as a function of the bubble radius and liquid parameters, has been proposed : $\text{We}_b = \text{Bo}_b^{-1/2} f(\text{Mo})$ where the Weber number $\text{We}_b= \rho V_\text{tip}^2 R_b/\gamma$ compares the effect of inertia and capillarity on the jet dynamics, $V_\text{tip}$ being the jet tip velocity as the jet passes the free surface level. 
\begin{figure}[htb]
\centering
\includegraphics[width=1\hsize]{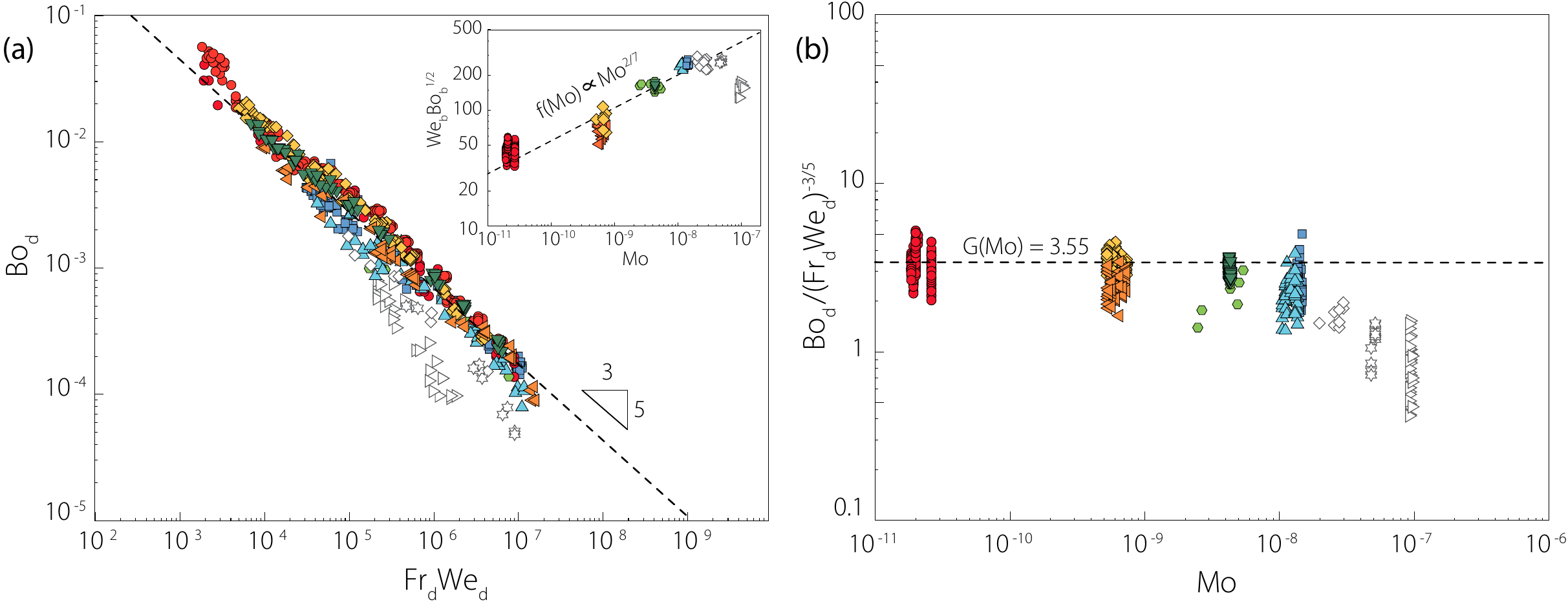}
\caption{(a) Drop Bond number as a function of the product of Froud and Weber number  $\text{Fr}_d\text{We}_d = \rho V_\text{tip}^4/ \gamma g$  for various values of the Morton number. For Mo $\lesssim 10^{-8}$, all the data, plotted with closed symbols, collapse on a single curve following the trend $\text{Bo}_d = \mathcal{C} \left(\text{Fr}_d\text{We}_d\right)^{-3/5}$ as shown by the dashed line. In the inset, the Fig.~3 (b) of Ghabache \textit{et al.} \cite{Ghabache2014} is plotted in a log-log plot, $f(\text{Mo}) = \mathcal{B} \text{Mo}^{2/7}$ fits reasonably well the data for $\text{Mo} \in [10^{-11} , 10^{-8}]$ as shown by the dashed line. (b)  $\text{Bo}_d / \left(\text{Fr}_d\text{We}_d\right)^{-3/5}$ as a function of Morton number. The dashed line fitting the data on the same range is a constant equals to $\mathcal{C}$. Data corresponding to $\text{Mo} \in [10^{-8} ,10^{-7}]$ plotted with open symbols leave the inviscid regime.  Above Mo $\simeq 10^{-7}$ no more drop can detach.
}
\label{Bod_vs_WeFr}
\end{figure}
Therefore, in the aim of decoupling the effects of bubble collapse and jet dynamics on the drop detachment, $\text{We}_b = \text{Bo}_b^{-1/2} f(\text{Mo})$ is combined with Eq.~\ref{BoLaw} in order to eliminate the bubble radius. This eventually yields the following scaling law relating the drop radius, the jet velocity and the liquid parameters : 
\beq
\text{Bo}_d = \left(\text{Fr}_d\text{We}_d\right)^{-3/5}  G(\text{Mo})
\label{BoLaw_2}
\eeq
where $\text{Fr}_d =  V_\text{tip}^2/gR_d$, leading to $\text{Fr}_d\text{We}_d = \rho V_\text{tip}^4/ \gamma g$ which
compares the effect of inertia upon capillarity and gravity on the jet dynamics, and is, in particular, independent of bubble radius and viscosity. In order to estimate $G(\text{Mo}) = H(\text{Mo}) f(\text{Mo})^{6/5}$, f(Mo) needs to be known. On the inset of Fig.~\ref{Bod_vs_WeFr},  $\text{We}_b \text{Bo}_b^{1/2}$ is plotted as a function of Mo (Fig.~3 (b) of Ghabache \textit{et al.} \cite{Ghabache2014}) in a log-log plot allowing us to determine f(Mo) by fitting the data in the same regime ($10^{-11} \lesssim \text{Mo}\lesssim10^{-8}$). The power law $f(\text{Mo})= \mathcal{B} \text{Mo}^{2/7}$, with $\mathcal{B}$ = 3.9~10$^4$, fits reasonably well the experimental data. Consequently, $G(\text{Mo}) = \mathcal{A} \text{Mo}^{-1/3}(\mathcal{B} \text{Mo}^{2/7})^{6/5}= \mathcal{C} \text{Mo}^{1/105} \sim \mathcal{C} $, with $\mathcal{C} = \mathcal{A} \mathcal{B}^{6/5} =3.55$.
This signifies that viscosity is removed from the scaling law relating the drop radius, the jet velocity and the liquid parameters, leading to,  for Mo $\lesssim 10^{-8}$ :
\beq
\text{Bo}_d = \mathcal{C} \left(\text{Fr}_d\text{We}_d\right)^{-3/5}.
\label{BoLaw_3}
\eeq
On the Fig.~\ref{Bod_vs_WeFr} (a) the drop Bond number Bo$_d$ is, therefore, plotted as a function of $\text{Fr}_d\text{We}_d$ and we observe an excellent collapse of all the experimental data represented with closed symbols, confirming Eq.~\ref{BoLaw_3}. Figure.~\ref{Bod_vs_WeFr} (b) presents $\text{Bo}_d / \left(\text{Fr}_d\text{We}_d\right)^{-3/5}$ as a function of Morton number and confirms that the drop Bond number is independent of viscosity. This inviscid behaviour stops at Mo $\simeq 10^{-8}$, viscosity playing a role for open symbols, between $10^{-8}$ and  $10^{-7}$  (corresponding to $\mu \sim$ 5 and 7mPa.s for water glycerol mixtures). Above Mo $\simeq 10^{-7}$ no more drop can detach.


Equation~\ref{BoLaw_3}, is therefore more robust, with less scattering than Eq.~\ref{BoLaw}. Furthermore, it demonstrates that viscosity is negligible in the detachment and drop size selection process. This result was predictable as the Ohnesorge number is always lower than one, but is in apparent contradiction with Figs.~\ref{fig:sequence_jet}, \ref{Bod_vs_Bob} and Eq.~\ref{BoLaw} where the drop radius clearly depends on the liquid viscosity.
But this influence of viscosity on drop size is only through the jet's formation as a memory of the bubble collapse. Indeed, as a bubble collapses, capillary waves focuses at the bottom of the cavity giving birth to the jet. Increasing the liquid viscosity changes the wave focusing, shooting up faster and thinner jet \cite{Ghabache2014}, therefore producing smaller droplets. 
In Eq.~\ref{BoLaw_3}, this shaping effect is then entirely contained through $V_{\text{tip}}$ and viscosity can disappear, shedding light on the inviscid behavior of the drop detachment mechanism.
Finally, the Bond number of the drop seems to be only selected by a competition between the given inertia, which makes the jet rising and stretching, and the duet gravity-capillarity which pulls on the jet tip so as to form a blob, initiating an end-pinching mechanism and consequently releasing a drop. While the influence of capillarity is obvious in this blob formation, the one of gravity can be more surprising. However, at the height the drop is detached, the gravity can already play a role. Indeed, the Froude number built on the drop detachment height $h_\text{det}$, Fr$h_\text{det} = V_\text{tip}^2/gh_\text{det}$, equals to O(1) for top drops projected by the largest bubbles.






As a conclusion, in this paper, we provide experimentally two different scaling laws giving the top jet drop radius as a function of the liquid parameters and the mother bubble radius in Eq.~\ref{BoLaw} or the jet velocity in  Eq.~\ref{BoLaw_3}. These results induce various outcomes.
The size distribution of the top jet drop aerosol can easily be computed as long as we know the bubble size distribution, which is the case in ocean for example \cite{Deane2002}. Note that the top jet drop plays a crucial role in terms of chemical exchange and evaporation, as it is usually bigger and faster than the followers \cite{Ghabache2016}. These results also apply to slightly viscous liquids (up to Mo $\sim 10^{-8}$) like champagne or sparkling wine for example. 
Furthermore, we untangled experimentally the intricate roles of viscosity, gravity and surface tension in the \textit{end-pinching} mechanism : typically liquid viscosity does not play any role in the drop detachment (Mo $\lesssim 10^{-8}$), contrary to the duet gravity-capillarity which initiates the drop detachment by pulling on the jet tip. Our results probably do not apply to other inertial stretched jet than those created by bubble bursting, as the intrinsic jet shape and size are hidden in the scaling law. But these results would need to be compared to the top breakup of other kind of stretched jets (cavity collapse after impact, bubble pinch-off etc.).

The Direction G\'en\'erale de l'Armement (DGA) is acknowledged for its financial support. 


\bibliographystyle{unsrt} 

%
%

\end{document}